\documentclass[aps,prd,nofootinbib,twocolumn,showpacs,superscriptaddress]{revtex4-1}

\pdfoutput=1
\usepackage {natbib}
\usepackage {graphicx}
\usepackage {hyperref}
\usepackage {color}
\usepackage{textcase}
\usepackage{amsmath, amsthm, amsfonts}
\usepackage{enumerate}
\usepackage{soul}
\usepackage[normalem]{ulem}

\newcommand{\ee}[1]{\!\times\!10^{#1}}

\usepackage[usenames,dvipsnames]{xcolor}

\newcommand{\Gauss}{\mathrm{\MakeUppercase{G}}}
\newcommand{\Signal}{{\mathrm{\MakeUppercase{S}}}}
\newcommand{\Line}{{\mathrm{\MakeUppercase{L}}}}
\newcommand{\Transient}{{\mathrm{t\MakeUppercase{L}}}}

\newcommand{\NoisetL}{{\Gauss\Line\Transient}}

\newcommand{\BSNtsc}{{\hat\beta}_{{\Signal/\NoisetL}}}	

\begin{document}

\title{Optimising the choice of analysis method for all-sky searches for continuous gravitational waves with Einstein@Home}

\author{Sin\'{e}ad Walsh}
\affiliation{Department of Physics, University of Wisconsin, Milwaukee, WI 53201, USA}
\affiliation{Max Planck Institute for Gravitational Physics (Albert Einstein Institute), D-30167 Hannover, Germany}
\author{Karl Wette}
\affiliation{ARC Centre of Excellence for Gravitational Wave Discovery (OzGrav) and Centre for Gravitational Physics, Research School of Physics and Engineering, The Australian National University, ACT 0200, Australia}
\affiliation{Max Planck Institute for Gravitational Physics (Albert Einstein Institute), D-30167 Hannover, Germany}
\author{Maria Alessandra Papa}
\affiliation{Max Planck Institute for Gravitational Physics (Albert Einstein Institute), D-30167 Hannover, Germany}
\affiliation{Department of Physics, University of Wisconsin, Milwaukee, WI 53201, USA}
\author{Reinhard Prix}
\affiliation{Max Planck Institute for Gravitational Physics (Albert Einstein Institute), D-30167 Hannover, Germany}

\begin{abstract}
Rapidly rotating neutron stars are promising sources of continuous gravitational wave radiation for the LIGO and Virgo interferometers. The majority of neutron stars in our galaxy have not been identified with electromagnetic observations. All-sky searches for isolated neutron stars offer the potential to detect gravitational waves from these unidentified sources. The parameter space of these blind all-sky searches, which also cover a large range of frequencies and frequency derivatives, presents a significant computational challenge. Various methods have been designed to perform these searches within the limits of available computing resources. Recently, a search method called Weave has been proposed to achieve template placement with a minimal number of templates. We employ a mock data challenge to assess the ability of this method to recover signals in an all-sky search for unknown neutron stars, and compare its search sensitivity with that of the global correlation transform method, which has been used for all-sky searches with the Einstein@Home volunteer computing project for a number of years. We find that the Weave method is 14\% more sensitive than GCT for an all-sky search on Einstein@Home, with a sensitivity depth of $57.9\pm0.6$ $\frac{1}{\sqrt{\mathrm{Hz}}}$ at 90\% detection efficiency, compared to $50.8^{+0.7}_{-1.1}$ $\frac{1}{\sqrt{\mathrm{Hz}}}$ for GCT. This corresponds to a 50\% increase in the volume of sky where we are sensitive with the Weave search. We also find that the Weave search recovers candidates closer to the true signal position. In the all-sky search studied here the improvement in candidate localisation would lead to a factor of $70$ reduction in the computing cost required to follow up the same number of candidates. We assess the feasability of deploying the search on Einstein@Home, and find that Weave requires significantly more memory than is typically available on a volunteer computer. We conclude that, while global correlation transform method remains the best choice for deployment on Einstein@Home due to its lower memory requirements, Weave presents significant advantages for the subsequent hierarchical follow-up searches of interesting candidates.

\end{abstract}

\maketitle

\section{Introduction}
\label{sec:introduction}

Continuous gravitational waves (CWs) from isolated neutron stars (NSs) are a potential source of detectable gravitational waves. CW radiation is emitted by rotating NSs with non-axisymmetric deformations. The signal is expected to be relatively stable over many years. While the amplitude of CW signals is expected to be small, the continuous nature of the signal allows us to integrate the signal over large time spans of data to distinguish it from noise.

Broad-band all-sky searches cover the whole sky over a broad range of frequency and frequency derivative in order to detect CW radiation from unknown NSs. All-sky searches in the Laser Interferometer Gravitational-Wave Observatory (LIGO) \cite{LIGO_1,LIGO_2} and Virgo \cite{Virgo_1,Virgo_2} data have so far not resulted in detection. Instead upper limits have been placed on the amplitude of CWs from isolated NSs \cite{EaH_O1,EaH_S6, MultiPipeline_O1, MultiPipeline_O1_fullband, S6FU}. The advanced detectors, which began operation in 2015, will eventually have a sensitivity to these weak signals over an order of magnitude better than that of the previous generation, with the largest gains at frequencies below 100\,Hz. 

The most sensitive search for CW signals is performed with a fully coherent integration over a large timespan, months to years, of data. The computational power required for the integration increases rapidly with the observation time of the data. When searching for CW signals over a broad frequency and spindown range, and over the whole sky, a fully coherent search quickly becomes computationally unfeasible \cite{VirgoCompCost,Brady}. This is the motivation for semi-coherent search methods. The data is split into shorter segments which are analysed coherently. Then, for each point in the parameter space $(f_0,\dot{f},\alpha,\delta)$, the coherent results in each segment are combined incoherently to determine the detection statistic over the entire set of data. For limited available computing power, these semi-coherent search methods achieve higher sensitivity than could be achieved with a fully coherent search \cite{Brady}.

Einstein@Home is a volunteer computing project where members of the public donate their idle computing power to the search for long lasting astrophysical signal \cite{EaH_url}.  The donated computing power allows for broader and more sensitive searches than previously possible \cite{EaH_O1,EaH_S6}. The Einstein@Home GW search that we consider here is described further in \cite{EaH_S6}. 

Some of the first semi-coherent searches used the Hough transform method \cite{Hough1,Hough2}, which sums weighted binary counts in each segment, depending on whether the normalised power in that segment exceeds a certain threshold. The parameter space used to sum the coherent results across segments is refined in both sky and spindown.  This method was applied to all-sky searches with the Einstein@Home computing project \cite{EaH_url} on LIGO S5 data \cite{EaH_S5}. The global correlation transform method (GCT) \cite{GCT_method} was later developed and claimed to be more sensitive at fixed computing cost, without the need for sky refinement in the incoherent part of the search. This search method was used in the Einstein@Home all-sky searches in LIGO S6 and O1 data \cite{EaH_S6,EaH_O1}. The GCT method is able to use arbitrarily long coherent segment lengths, however, its sensitivity cannot be reliably estimated analytically, instead extensive injection-and-recovery studies are needed to design an all-sky search using this method \cite{S6FU}. 

The purpose of the study presented in this article is to assess a recently-developed semi-coherent search method to detect CWs from isolated neutron stars. A principal claim of this method is that it achieves template placement with a minimal number of templates required to achieve a desired maximum mismatch on the semi-coherent and coherent search grid in all-sky searches \cite{Weave1,Weave2,Weave3,Weave4}. This would make it a milestone tool, achieving the highest search sensitivity possible at fixed computing cost, with increases in sensitivity only possible if the code can be made faster. An additional, and significant, benefit of having optimal template placement is that we can predict the sensitivity without the need for extensive injections-and-recovery simulations. The implementation of this semi-coherent search method is described in detail in \cite{WeaveImp2018}, and is hereafter referred to as Weave.

We want to examine the detection efficiency of this search method in an all-sky search for isolated neutron stars with the Einstein@Home project. This type of search is currently performed with the GCT method. With the GCT search method, the optimal search setup (i.e. the optimal combination of search parameters) at fixed computational cost is determined by running Monte-Carlo injection-and-recovery studies of many search setups \cite{S6FU}.

With the Weave search method, it should be possible to
semi-analytically determine optimal search setups, without the need
for extensive Monte-Carlo injection-and-recovery studies.
This would greatly simplify the process of setting up an all-sky search.
This is because with Weave we can reliably predict the
distribution of relative SNR-loss (in the following referred to as
\emph{mismatch}) of the Weave template bank for any given
search-setup (see \cite{WeaveImp2018,Weave4} for details, and
\cite{octapps} for an implemenation of the mismatch predictor).
This is a key input required for accurate sensitivity
\emph{estimation} (as described in \cite{Sensitivity2018}),
as well as for the semi-analytic sensitivity \emph{optimization}
method developed in \cite{Prix2012,octapps}.
This method has since been extended to take advantage of
empirical non-linear mismatch predictions (as given in \cite{Weave4}),
and work is ongoing to apply this framework to the Weave search
method. The results of this will be reported in future work.

The first objective of this paper is to compare the achievable detection efficiency of the Weave search method with that of existing semi-coherent methods, especially the GCT search method. The second objective of this paper is to determine if the Weave search method can be applied within the technical limitations of searches with the Einstein@Home project. 

This comparison is first made using the same data as in the mock data challenge (MDC), which was used to compare five all-sky search methods for the detection of gravitational waves from unknown neutron stars \cite{MDC}. The comparison is repeated in simulated detector data (Gaussian noise) with more fake signals, to allow for a more precise comparison with the GCT search. This comparison is made for the standard CW signal model described in Sections \ref{sec:signal} and \ref{sec:MDC}.

An overview of the search method is presented in Section \ref{sec:method}. The MDC is described briefly in Section \ref{sec:MDC}; and the choice of optimal setup for the Weave search is described in Section \ref{sec:setup}. Section \ref{sec:results} describes how the methods are compared and the results of the comparison.

\section{The signal}
\label{sec:signal}

Gravitational waves (GWs) emitted from non-axisymmetric NSs are typically described by a signal which remains relatively stable over years of observation \cite{Fstat}. The strain amplitude of the GW is proportional to the ellipticity, $\varepsilon$, defined as 
\begin{equation}
\varepsilon = \frac{|I_{xx} - I_{yy}|}{I_{zz}},
\end{equation}
where $I_{zz}$ is the principal moment of inertia of the star, and $I_{xx}$ and $I_{yy}$ are the moments of inertia about the axes. The strain amplitude of the GW at the detector, assuming a rigidly rotating triaxial body, is then given by 
\begin{equation}
h_0 = \frac{4\pi^2G}{c^4}\frac{I_{zz}f^2\varepsilon}{d},
\end{equation}
where $f$ is the frequency of the GW, $G$ is the Gravitational constant, $c$ is the speed of light, and $d$ is the distance to the NS. For a star steadily rotating around a principal axis of inertia, the frequency of the GW is at twice the rotational frequency of the NS. The frequency evolves over time as energy is lost due to various dissipation mechanisms, including GW emission. The first time derivative of the frequency, $\dot{f}$, is referred to as spindown.

The signal model is described by eight parameters, four phase evolution parameters $\lambda = (f_0,\dot{f},\alpha,\delta)$ and four amplitude parameters $(h_0,\iota,\psi,\phi_0)$, where $\iota$ is the inclination angle between the line of sight to the NS and its rotation axis, $\psi$ is the polarisation angle and $\phi_0$ is the initial phase of the signal at a reference time $\tau_0$. 

The parameter $f_0$ is the GW frequency at $\tau_0$. The GW frequency in the Solar System Barycenter (SSB) at time $\tau$ is given by

\begin{equation}
f_\mathrm{SSB}(\tau) = f_0 + \dot{f}(\tau-\tau_0),
\label{eq:fssb}
\end{equation}
assuming the second time derivative of the frequency, $\ddot{f}$, is negligible. The transformation between the time $t$ at which a certain wavefront arrives at a detector and the time $\tau$ at which the same wavefront arrives at the SSB is

\begin{equation}
\tau(t) = t + \frac{\bold{r}(t) \cdot \bold{n}}{c} + \Delta_{E\odot} - \Delta_{S\odot}
\label{eq:track}
\end{equation}

where $\bold{r}(t)$ is the position vector of the detector in the SSB frame, $\bold{n}$ is the unit-vector pointing towards the sky location of the source, and $\Delta_{E\odot}$ and $\Delta_{S\odot}$ are the relativistic Einstein and Shapiro time delays respectively. Ignoring the relativistic corrections, the instantaneous frequency $f(t)$ of a continuous GW signal as observed at the detector is related to the frequency $f(\tau)$ at the SSB by the relation:

\begin{equation}
f(t) = f_\mathrm{SSB}(\tau) + f_\mathrm{SSB}(\tau) ( \frac{\bold{v}(\tau) \cdot \bold{n}}{c} )
\label{eq:track}
\end{equation}

where $\bold{v}(\tau)$ is the detector velocity with respect to the SSB frame. The frequency measured in the detector, $f(t)$, is shifted due to Doppler effect from the motion of the Earth around the Sun and by the rotation of the Earth.

In a blind all-sky search there is the potential for the detection of signals produced by different emission mechanisms (e.g.\ $r$-modes \cite{Owen_f2}). The ability of the all-sky search methods to recover such signals is not examined in this study. Here we assume the signal follows the model described above.\\

\section{The Weave method}
\label{sec:method}

The search begins with many months of data being split into segments of a few days each. For the coherent analysis, the multi-detector $\mathcal{F}$-statistic \cite{Fstat,Fstatmulti} is computed for each segment and for each parameter space point on the coherent grid. The coherent grid is determined by the Weave method, which is founded on the use of the reduced supersky metric, introduced in Section B IV of \cite{Weave1}. The reduced supersky metric is a close approximation to the phase metric, and expressed in a convenient coordinate system with respect to which the metric is constant and numerically well-conditioned.

An average $2\mathcal{F}$-statistic, $2\mathcal{\overline{F}}$, is calculated by summing $2\mathcal{F}$ values for a single template across the segments. (A template waveform is defined by the four phase evolution parameters $\lambda = (f_0,\dot{f},\alpha,\delta)$.) A second grid in parameter space is set up, the so-called semi-coherent grid. This grid is finer than the coherent grid in spindown and sky \cite{Weave3}. The detection statistic at every point in the semi-coherent grid is estimated by summing detection statistic values at the closest points on the coherent grid. Weave uses a metric based method described in \cite{Weave3} to determine the closest point on the coherent grid.

\section{The mock data challenge}
\label{sec:MDC}

In \cite{MDC}, a MDC was employed to empirically compare the performance of different all-sky search methods. This was done by simulating the detector response to CW signals in data from the S6 LIGO science run \cite{S6data}, with fake signals at frequencies ranging from $40$ to $1550\,$Hz. Each of the search pipelines then performed a search over the data to assess its ability to recover the signal. \\

The sensitivity is measured in terms of the smallest signal amplitude  detectable with a given efficiency $h_{\text{smallest-detectable}}$. The smallest detectable amplitude scales linearly with the noise $\sqrt{S_{h}}$ so the ratio of the smallest detectable amplitude to the $\sqrt{S_{h}}$ is roughly constant for a given search across the different frequencies. The smaller the ratio, the higher the sensitivity of the search. For this reason the notion of sensitivity depth $\cal{D}$ was introduced \cite{GalacticCenter}: ${\cal{D}}:= {\sqrt{S_h}\over {h_{\text{smallest-detectable}}}}$. The higher the sensitivity depth, the more sensitive a search is.
When comparing the Weave and GCT searches we quote here their sensitivity depths at 90\% detection efficiency.

Note that in \cite{MDC} the signal strength is normalised by $S_h$ defined as the harmonic mean over both detectors of the harmonic sum power spectral density (PSD) of the data, at the frequency of the signal. Instead here we define $S_h$ as the harmonic mean over both detectors of the harmonic mean PSD of the data, at the frequency of the signal. One can approximately\footnote{assuming roughly equal number of SFTs from each detector} convert the former into sensitivity depth by multiplying by $\sqrt{N_\mathrm{det}} \approx 1.4$, as 2 detectors were used. A discussion of why the harmonic mean over per-detector PSDs should be used in the calculation of $S_h$ is included in \cite{Sensitivity2018}. 
 
In the MDC of \cite{MDC}, different search configurations were assumed for the Einstein@Home search with GCT in different frequency ranges, namely from 40 to 500\,Hz, 500 to 1000\,Hz and 1000 to 2000\,Hz. Here we perform the same MDC with Weave only in the frequency range of 40 to 500\,Hz, which is particularly interesting because it covers the most sensitive range of the detectors.  

In Section \ref{sec:MDC_efficiency} we assess the performance of the Weave search method with this MDC data and compare it to the sensitivity of the searches in Figure 3 of \cite{MDC}. 

We are most interested in comparing the Weave search method with the GCT method which was used by the Einstein@Home search in the MDC. However, the difference in sensitivity between the Weave search and the GCT search is less than the uncertainty on the efficiency measurement in the MDC (This is shown in Section \ref{sec:MDC_efficiency}, Figure \ref{fig:MDC}). Therefore, we generate a more extensive data set so we can make a more precise comparison between the Weave and GCT search methods. 

The fake signals in this new set are created in almost the same way as the fake signals used in the MDC, described in Section IV A of \cite{MDC}. Some differences are, we only generate injections in the frequency range of 40 to 500\,Hz, and we do not include injections with positive $\dot{f}$ or with non-zero $\ddot{f}$, both of which were found to have no effect on the measured efficiency.

In the MDC, the $h_0$ of fake signals was obtained by randomly drawing the SNR from a uniform distribution. However, it was shown in \cite{KarlSNRbias} that, at fixed signal strength $h_0$, a uniform SNR distribution implies an unphysical distribution of $\iota$ which is biased towards higher absolute values of cos$\,\iota$. The surplus of higher absolute values of cos$\,\iota$ results in an overestimate of the CW search sensitivity by $\approx 30\%$ \cite{KarlSNRbias} at fixed signal strength. When generating new fake signals, we instead obtain the $h_0$ by randomly drawing from a uniform distribution of normalised signal strength. 

The range of signal strength was chosen to bracket the 90\% detection efficiency for both GCT and Weave, because search sensitivities are typically compared at high detection efficiencies \cite{MDC}. These are comparable to the upper limits values expected from the searches, in case of a null result.

For the MDC the fake signals were injected into 15 months of LIGO S6 data. The GCT search was performed using only the last nine months of this data set. Here, we do not use LIGO S6 data. Instead, we generate nine months of Gaussian noise data with the same duty cycle as the LIGO S6 data. For this comparison between Weave and GCT we are interested in the sensitivity in well behaved data. By generating fake data we avoid being affected by unclassified detector disturbances in S6 data.

\section{Choosing the search setup}
\label{sec:setup}

The search parameters and selection criteria to use in an Einstein@Home all-sky search depend on the parameter space to be covered and the computational budget. The GCT search set-up is the same as the one examined in the MDC \cite{MDC}. The Weave search is determined through an optimisation process, aimed at achieving the maximum sensitivity depth at fixed computational cost over the same search parameter space as for the GCT search. In particular we considered an all-sky search covering 40 to 500\,Hz and 2.9e-9\,Hz/s, and 6 months run-time on Einstein@Home \footnote{This assumes the search is run continuously on 8000 'fast' CPUs. In reality, there are more Einstein@Home CPUs, many slower than the CPUs assumed for the runtime estimate. At the time of the MDC \cite{MDC}, 8000 'fast' CPUs was determined to be an reasonable estimate for the speed of an Einstein@Home search.}. We also set a threshold on the recovered detection statistic such that we would expect a fixed number of candidates above threshold from false alarms in random noise in an all-sky search. This threshold and number of candidates is established, in Sections \ref{sec:effective_ntemplates} and \ref{sec:parameter_estimation}, to be such that the time taken to follow-up candidates from the Weave or GCT search would be equivalent. 

At a given coherent segment length, the Weave search takes two input parameters, the maximum metric mismatch for the coherent search grid, $\tilde{\mu}_{\mathrm{max}}$, and for the semicoherent search grid, $\hat{\mu}_{\mathrm{max}}$. These will determine the search grids. We find the optimal search setup by running Monte-Carlo injection-and-recovery studies of many search setups, as done for the GCT method in \cite{S6FU}. We then repeat for different coherent segment lengths. We compare the detection efficiencies at a threshold corresponding to a fixed false alarm rate, and choose the most sensitive setup.

Einstein@Home searches are designed to run for a certain number of months on the Einstein@Home computing project. Therefore, we need to be able to predict the total runtime of different search setups on Einstein@Home. The runtime model for the Weave search method is described in Section IV C of \cite{WeaveImp2018}, and its implementation is documented in \cite{octapps}.

The runtime prediction relies on having an estimate of the total number of search templates. We also need to know the number of templates to determine the false alarm rate for different thresholds on the detection statistic. 

\subsection{Determining the number of templates}
\label{sec:nsearch_templates}

Knowing the total number of templates searched is necessary for two reasons: 1) to determine the runtime of a search setup and  2) to predict the false alarm rate in Gaussian noise at different $2\mathcal{\overline{F}}$ thresholds, as described in \cite{EaH_S6}. 

A model to estimate the number of templates in a Weave search was not available when the study in this paper began. Instead, the number of templates was obtained by running the Weave search code over a small region of parameter space and counting the number of templates generated. The number of templates on the full search volume was then determined by scaling arguments to be $7\ee{16}$ templates on the coherent grid and $1\ee{18}$ semi-coherent templates. The GCT all-sky search would have $1\ee{15}$ templates on the coherent grid, and $2\ee{17}$ semi-coherent templates.

A model to estimate the number of templates in a Weave search has since been developed, and is described in Section IV B of \cite{WeaveImp2018}. The model has been implemented in a script which is publicly available, and given in \cite{WeaveImp2018}. The number of templates estimated with this script is expected to be accurate within a factor of a few. Such a deviation in the number of templates could result in a factor of a few difference in the estimated runtime for the search. This uncertainty is too large for a search on Einstein@Home; a search which is designed to run for six months in Einstein@Home should not run for a year or more. 

This script estimates that there are $6.1\ee{16}$ templates on the coherent grid and $2.8\ee{18}$ semi-coherent templates. This results in an approximate runtime for the all-sky search of 13.5 months on Einstein@Home. This is twice the runtime estimated by running the Weave search over a small region of parameter space and extrapolating from the number of templates counted.  

The number of templates required to cover any parameter space includes some padding of the parameter-space boundaries. When the total template count is extrapolated, the padding is also extrapolated, but the padding doesn't extrapolate in the same way as the bulk template count. When the script from \cite{WeaveImp2018} estimates the number of templates for the entire search over the whole sky and frequency range, it does not assume the same padding which would apply when the parameter space is split into smaller regions. Einstein@Home searches are split into smaller regions of parameter space, which are searched separately. To estimate the number of templates in an Einstein@Home search, the script which estimates the number of templates should be run separately for each sub-region in parameter space. When the number of templates are estimated in this way, there are $2.3\ee{16}$ templates on the coherent grid and $1\ee{18}$ semi-coherent templates. This results in an approximate runtime for the all-sky search of 4.9 months on Einstein@Home. 

This estimate of 4.9 months is reasonably similar to the runtime of 6 months estimated by calculating the number of templates from scaling arguments, and is significantly smaller than the runtime estimate of 13.5 months obtained by estimating the number of search templates for the entire search as if it was a single search. However, for this search spanning $40$ to $500$\,Hz, running this script in series for each $0.05$\,Hz band took more than a week. Therefore, the script to estimate the number of templates from \cite{WeaveImp2018} is not suitable for an Einstein@Home search. However, it is suitable to estimate the number of templates for all-sky searches which are not split into many sub-regions in parameter space.

\subsection{The optimal search setup}

The best Weave setup found for an all-sky search from 40 to 500\,Hz, over nine months of data, which runs for six months on Einstein@Home, is given in Table \ref{tab:setup_MCs}. This setup is found using Monte-Carlo injection-and-recovery studies as described in Section \ref{sec:setup}. The GCT search parameters used in the MDC are shown for reference in Table \ref{tab:EaH_setup}. The number of semi-coherent templates in each dimension of parameter space are shown for a nominal region of parameters space in Table \ref{tab:ntemplates}.

\begin{figure}[htb!]
  \includegraphics[width=3.2in]{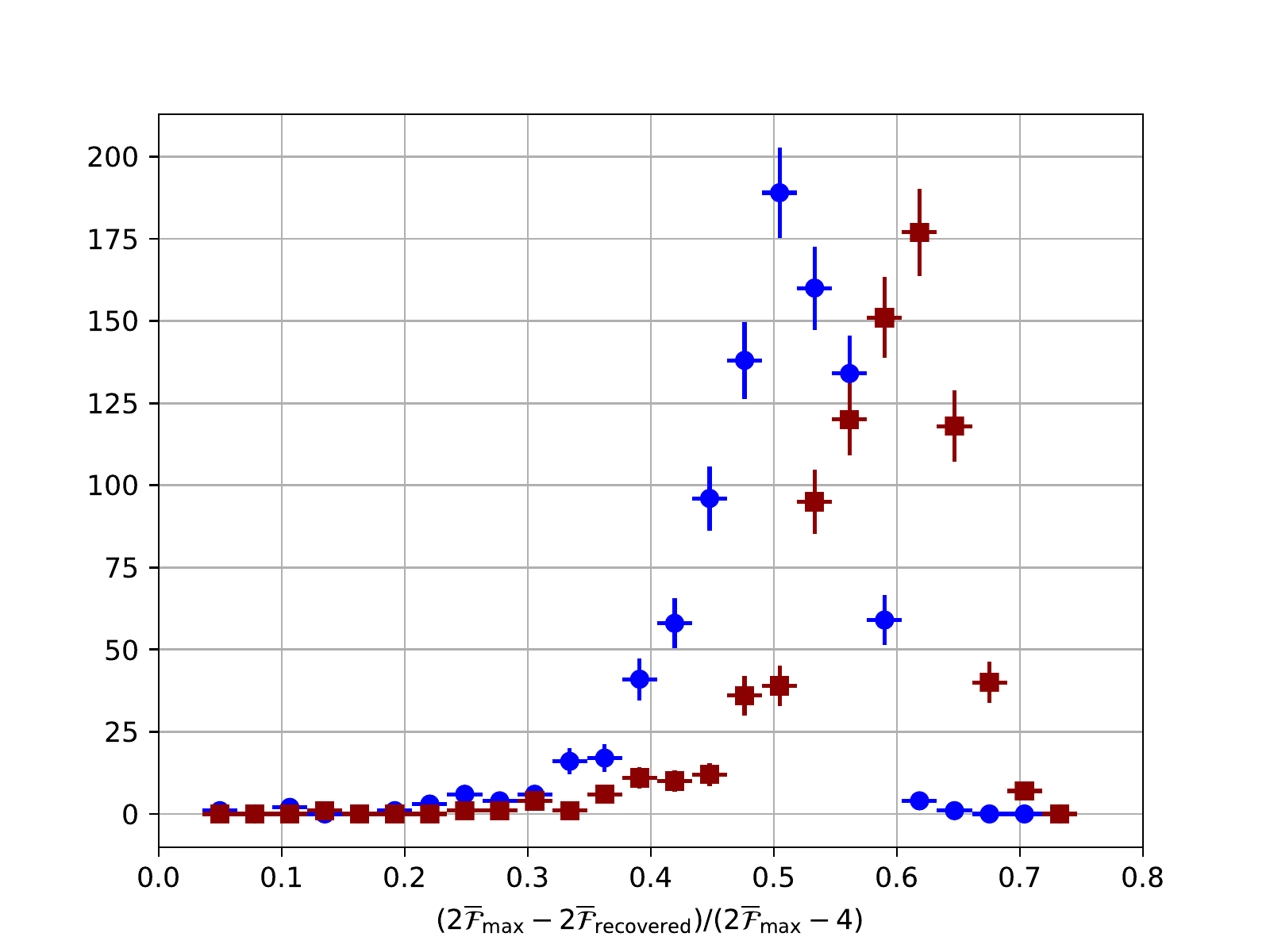}\\

\caption{\label{fig:mismatch} Histogram of the mismatch between the maximum $2\mathcal{\overline{F}}$, the $2\mathcal{\overline{F}}$ of a search template at the exact signal parameters, and the loudest recovered $2\mathcal{\overline{F}}$ from a GCT search (red) and a Weave search (blue).}
\end{figure}

In Figure \ref{fig:mismatch} we plot the mismatch between the loudest recovered $2\mathcal{F}$ and the maximum possible $2\mathcal{F}$ i.e the $2\mathcal{F}$ we would recover with a search template exactly at the signal parameters. This mismatch indicates how much of the signal SNR we are losing due to the offset between the true signal parameters and the parameters of the closest search template. The Weave search has a mean mismatch of 0.49. The GCT search has a mean mismatch of 0.59. 

\begin{table}[h!]
\begin{tabular}{|c|c|}
\hline
$\mathrm{T_{coh}}$ (h) & 60 \\
$\mathrm{N_{segments}}$ & 100 \\
$\tilde{\mu}_{\mathrm{max}}$ & 0.2 \\
$\hat{\mu}_{\mathrm{max}}$& 20 \\
\hline
\end{tabular}
\caption{Search parameters for the Weave MDC search, as determined by Monte-Carlo injection-and-recovery studies. $\mathrm{T_{coh}}$ is the coherent segment length in hours. $\tilde{\mu}_{\mathrm{max}}$ is the maximum coherent mismatch. The frequency spacing on the coherent grid for this setup is $1.61\ee{-6}$ Hz. $\hat{\mu}_{\mathrm{max}}$ is the maximum semicoherent mismatch.}
\label{tab:setup_MCs}
\end{table}

\begin{table}[h!]
\begin{tabular}{|c|c|}
\hline
$\mathrm{T_{coh}}$ (h) & 60\\
$\mathrm{N_{segments}}$  & 90 \\
$\delta f$ (Hz) & $3.61\ee{-6}$\\
$\delta \dot{f}$ (Hz/s) & $1.16\ee{-10}$\\
sky factor & 0.01 \\
$\dot{f}$ refine & 230\\
\hline
\end{tabular}
\caption{Einstein@Home GCT MDC search parameters. The sky grid resolution is determined by the sky factor as shown in Equation 8 of \cite{MDC} The spindown resolution used on the fine grid, for the semi-coherent part of the search, is given by $\delta \dot{f}$ divided by the $\dot{f}$-refine value.}
\label{tab:EaH_setup}
\end{table}

\begin{table}[h!]
\begin{tabular}{|c|c|c|}
\hline
$\mathrm{N_{templates}}$&Weave&GCT\\
\hline
0.05 Hz & 31,055 & 13,850\\
2.9e-9 Hz/s & 5,750 & 1,204 \\
all-sky at 44Hz & 67,122 & 8,090 \\
\hline
\end{tabular}
\caption{The number of semi-coherent templates in each dimension of parameter space for a nominal region of parameter space: 0.05 Hz, $2.9\ee{9}$ Hz/s and the whole sky at 44 Hz for Weave and 50 Hz for GCT. The GCT sky grid templates are generated for every 10 Hz, so searches over frequencies from 40 to 50 Hz all use the 50 Hz sky grid. The number of sky grid templates in a Weave search scales with frequency.}
\label{tab:ntemplates}
\end{table}

\section{Comparison of methods}
\label{sec:results}

\subsection{Effective number of search templates}
\label{sec:effective_ntemplates}

The distribution of the average $2\mathcal{F}$ detection statistic, $2\mathcal{\overline{F}}$, in Gaussian noise is expected to follow a chi-squared distribution with $4N_{seg}$ degrees of freedom. From this we can estimate the number of false alarms above a $2\mathcal{\overline{F}}$ value for a search with N independent templates \cite{Noisechi2}. 

In Section \ref{sec:nsearch_templates} we calculate the number of templates, $N_{\mathrm{templates}}$, in the full all-sky search. Assuming these templates are independent, a $2\mathcal{\overline{F}}$ threshold of $6.1$ should result in 40 million false alarms in the all-sky search. However, if the search templates are not fully independent the effective number of independent templates is smaller. 

In this section we perform many searches over a small region of parameter space in Gaussian noise. For Weave the search range is 0.05 Hz, 1e-10 Hz/s, and a sky patch with 0.03 radians radius. For GCT the search range is 0.05 Hz, 3e-10 Hz/s, and a sky patch with a radius of 0.03 when projected onto the ecliptic plane. The Weave search has $\approx 3e+8$ templates and the GCT search has $\approx 5e+8$ templates.

We compare the distribution of loudest recovered $2\mathcal{\overline{F}}$, $2\mathcal{\overline{F}}_\mathrm{max}$, with what we expect if the search templates are independent. By fitting the expected $2\mathcal{\overline{F}}_\mathrm{max}$ distribution to the measured distribution, we can determine the effective number of templates in the search. \footnote{The shape of the $2\mathcal{\overline{F}}_\mathrm{max}$ distribution over correlated templates does not always agree with the $2\mathcal{\overline{F}}_\mathrm{max}$ distribution over $N_{eff}$ independent templates, see Appendix C of \cite{Sensitivity2018}}

\begin{figure}[htb!]
  \includegraphics[width=3.2in]{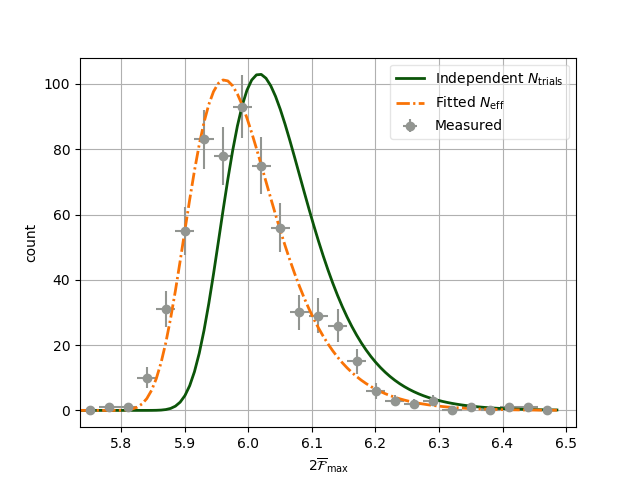}\\
  \includegraphics[width=3.2in]{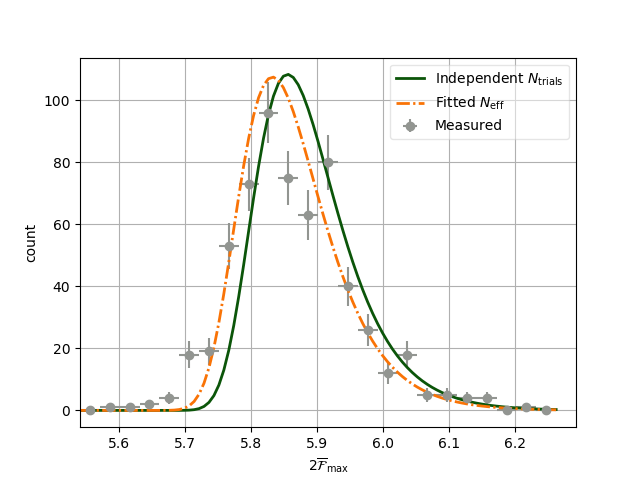}\\

\caption{\label{fig:GCT_Weave_noise} The distribution of recovered $2\mathcal{\overline{F}}_\mathrm{max}$ values in 600 repeated runs over Gaussian noise for GCT (top) and Weave (bottom). If the search templates were independent the measured $2\mathcal{\overline{F}}_\mathrm{max}$ (grey) would agree with expected $2\mathcal{\overline{F}}_\mathrm{max}$ for $N_{\mathrm{templates}}$ (green). The effective number of templates is obtained with a fit to the measured $2\mathcal{\overline{F}}_\mathrm{max}$, the results of the fit are shown in orange.} 
\end{figure}

The results are shown for the GCT search and for the Weave search in Figure \ref{fig:GCT_Weave_noise}. For the GCT search we measure an effective number of templates $N_{eff} =0.43 N_{\mathrm{templates}}$. For the Weave search $N_{eff} = 0.69 N_{\mathrm{templates}}$. 

In an all-sky search with GCT, with $N_{eff} =0.43 N_{\mathrm{templates}}$, we would expect 16 million false alarms in Gaussian noise with $2\mathcal{\overline{F}} > 6.17$, the threshold used for the MDC. In an all-sky search with Weave, with $N_{eff} = 0.69 N_{\mathrm{templates}}$, a threshold of $2\mathcal{\overline{F}} > 6.155$ would produce the same number of false alarms in Gaussian noise. 

\subsection{Memory usage}
\label{sec:memory}

When a search is run on the Einstein@Home computing project, the parameter space to search is split into cells, and each volunteer computer searches a cell corresponding to a specific work-unit (WU) of the global task. The cells are chosen such that a single WU will run for approximatly 8 hours on a volunteer computer. The loudest candidates recovered from each cell are then returned to the Einstein@Home servers.

While many modern computers have more than 10 GB of memory, Einstein@Home includes older computers with less memory. The searches run single-threaded, so typically one task is executed per available core, meaning the memory we can use is limited to what is available per CPU core. Therefore, to be able to run on Einstein@Home volunteer computers, we limit the memory a WU can use to 1 GB. Normally, the parameter space of an all-sky search is split into 0.05 Hz bands, and then the number of sky grid points is scaled to result in a search which runs for 8 hours. The GCT searches consume a managable amount of memory with such WUs. The Weave search given in Table \ref{tab:setup_MCs}, when split up to run for 8 hours over 0.05 Hz, reaches a memory consumption of 2-7 GB, depending on how you split the parameter space.

There are a number of ways to reduce the memory consumption for a WU. As an example, it is possible to stage, in sequence, multiple searches within a WU such that the total runtime of the WU remains close to 8 hours, but at any given time the memory consumption is smaller than if all the work was staged at the same time. This will results in more output files as each search staged will output a list of the most significant candidates, but this is something which can be handled by Einstein@Home. Alternatively, the Weave executable has an option to internally split the frequency and spindown parameter space of a WU into subsets, to limit the memory consumption. 

With both of the options for reducing the memory, there are limitations to how small you can make the parameter space. If you reduce the frequency range so there are fewer than $\approx 10k$ frequency bins in a single search, the runtime will increase as the resampling method for calculating the $\mathcal{F}$ becomes less efficient. The search runtime is also found to increase for small spindown search ranges and for small sky patch sizes. 

As a concrete example we attempt to determine the WU configuration which would work for an Einstein@Home search with the Weave search given in Table \ref{tab:setup_MCs}. However, we find that if the frequency band is less than 0.05 Hz the runtime increases, for a spindown range less than 2.9e-9 Hz/s the runtime increases, and for more than 500 sky patches at 150 Hz the runtime increases. At this limiting search configuration covering 0.05 Hz, 2.9e-9 Hz/s and 500 sky patches at 150 Hz, the memory consumption is 7 GB. If we decrease the WU parameter space to 0.04 Hz, 2.9e-9 Hz/s split internally in 10 partitions, and 500 sky patches at 150 Hz, the memory usage is 1 GB at a cost of a 50\% increase in the total search runtime. 

A semi-analytical model for predicting the memory usage is described in \cite{WeaveImp2018}. This can be used to design WUs within the memory limit constraints given by specific hardware. One of the parameters needed by the memory model is the maximum size of the Weave code cache, and this cannot easily be predicted. Instead, the user must run the Weave code in a special mode, which is faster than the normal mode, which measures the size of the cache without computing any $2\mathcal{F}$ values. This must be done for each WU configuration considered. This is a step in the deployment of a Weave search on Einstein@Home which is not necessary for a GCT search. On the other hand, memory footprints of a few GBs should pose no problem if using Weave on a dedicated super computing cluster. 

\subsection{The line-robust statistic}
\label{sec:linerobust}

The most recent all-sky search on Einstein@Home \cite{EaH_O1} has used the $\BSNtsc$ statistic as well as the $2\mathcal{\overline{F}}$ statistic. This is a line- and transient-robust statistic that tests the signal hypothesis against a noise model which, in addition to Gaussian noise, also includes single-detector continuous or transient spectral lines \cite{Keitel1,Keitel2}.

Including the $\BSNtsc$ statistic increases the runtime and memory usage of a search, as the additional statistics must be calculated for each search template. For the GCT and Weave searches considered here, including $\BSNtsc$ increases the runtime by $\approx 40\%$ and $\approx 250\%$, respectively.  This increase may be different for other setups. The additional computing cost for Weave would require a repeat of the optimisation procedure in Section \ref{sec:setup} with a runtime model which includes the extra statistic, which would ultimately lead to a lower efficiency than the current setup at fixed computing cost. 

For the setup considered here, the memory increase is negligible for GCT, which remains well below 1GB, whereas it increases by a factor of $\approx$ 3 for Weave. This memory increase can potentially be managed as described in Section \ref{sec:memory}, at the expense of a higher runtime.  

The increase in runtime and memory is due to the fact that the extra statistics are calculated for every point in parameter space. However, this statistic is only beneficial (compared to the $2\mathcal{\overline{F}}$ statistic) in regions of parameter space where the data is disturbed by detector artifacts, which was 10\% of the frequency bands in the S6 all-sky search \cite{EaH_S6}. By calculating the $\BSNtsc$ statistic only on the disturbed frequency bands it may be possible to apply the optimal Weave search setup within the computing budget while retaining the benefits of the $\BSNtsc$ statistic. 

Due to the memory limitations of Einstein@Home (Section \ref{sec:memory}) it would not be possible to run this postprocessing stage on Einstein@Home, instead it would be run on a computing cluster. Since the original search is designed to run for 6 months over 8000 CPU cores, if as little as 5\% of the frequency bands are disturbed the search would take a month to run continuously on 5000 cluster CPU cores. If 10-20\% of the frequency bands are disturbed, rerunning with the $\BSNtsc$-statistic becomes prohibitively expensive.

When we say a frequency band is disturbed, we mean that the list of the candidates with the largest $2\mathcal{\overline{F}}$ statistic in that frequency band is dominated or saturated by candidates due to a detector artefact. When we use the $\BSNtsc$-statistic in the search, a separate list of the candidates with the largest $\BSNtsc$-statistic is output. The advantage of including the $\BSNtsc$-statistic is when a frequency band which appears disturbed in the $2\mathcal{\overline{F}}$-statistic list is not disturbed in the $\BSNtsc$-statistic list, which means the candidates from the detector artefact have been suppressed and there is still the potential to recover a signal in this frequency band.

Even if 10\% of frequency bands are disturbed, only a fraction of these will be recovered by using the $\BSNtsc$-statistic. If this fraction is as low as 20\% (2\% of the total), we would not consider the exclusion of the $\BSNtsc$-statistic to be a noteworthy disadvantage of the Weave search. However, it is realistic to expect that 50\% of disturbed bands would be recovered by using the $\BSNtsc$-statistic. For the comparison in this paper we consider a search using only the $2\mathcal{\overline{F}}$ statistic.

\subsection{Results of the comparison}
\subsubsection{Signal parameter estimation}
\label{sec:parameter_estimation}

In an all-sky search, above-threshold candidates are passed through hierarchical refinement stages designed to exclude candidates from noise while retaining signal candidates \cite{S6FU}. In the MDC, the parameter space around the candidates in the first refinement stage is used to determine which candidates from the initial stage would result in a detection.

\begin{figure}[htb!]
  \includegraphics[width=3.2in]{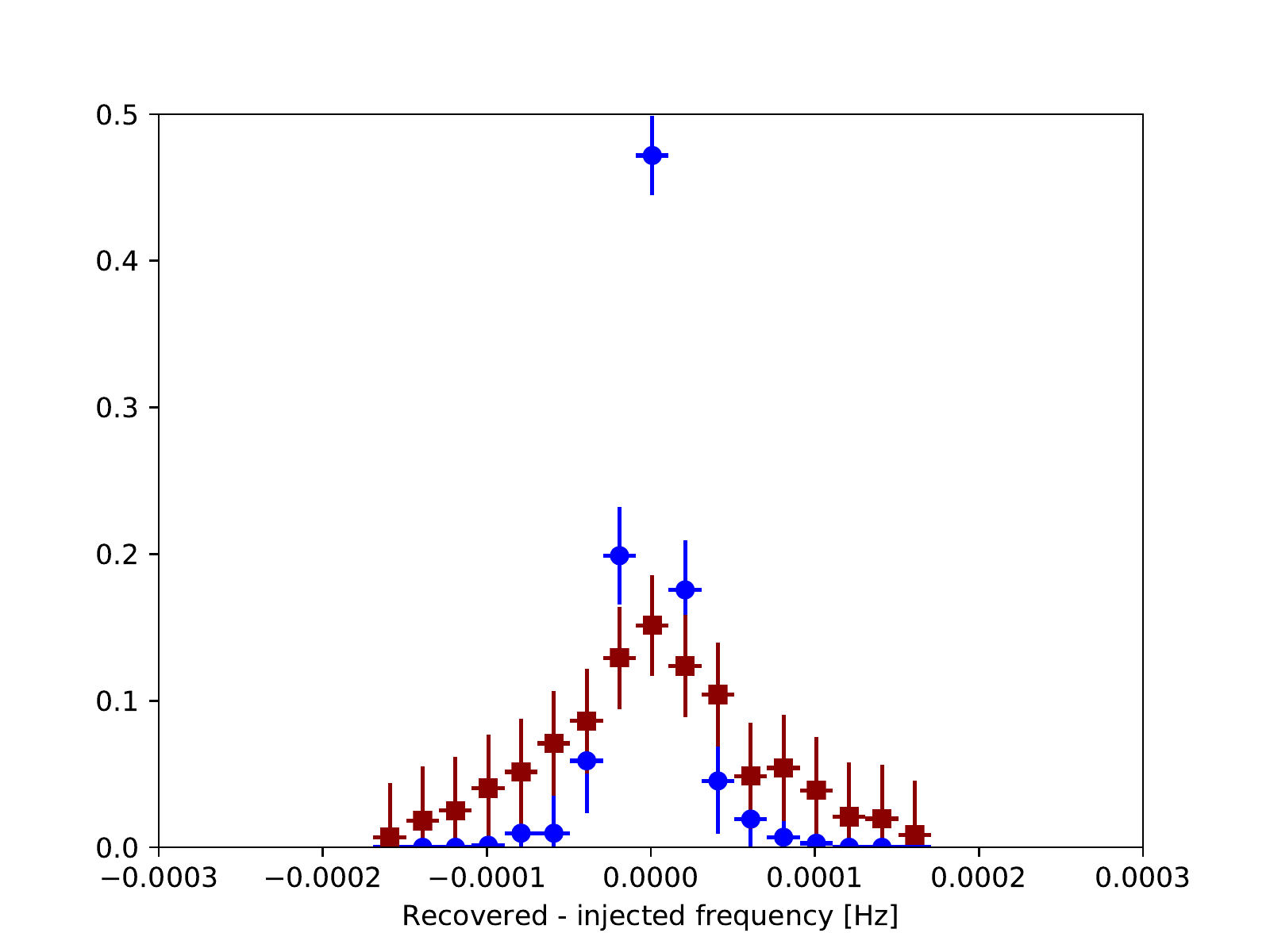}\\
  \includegraphics[width=3.2in]{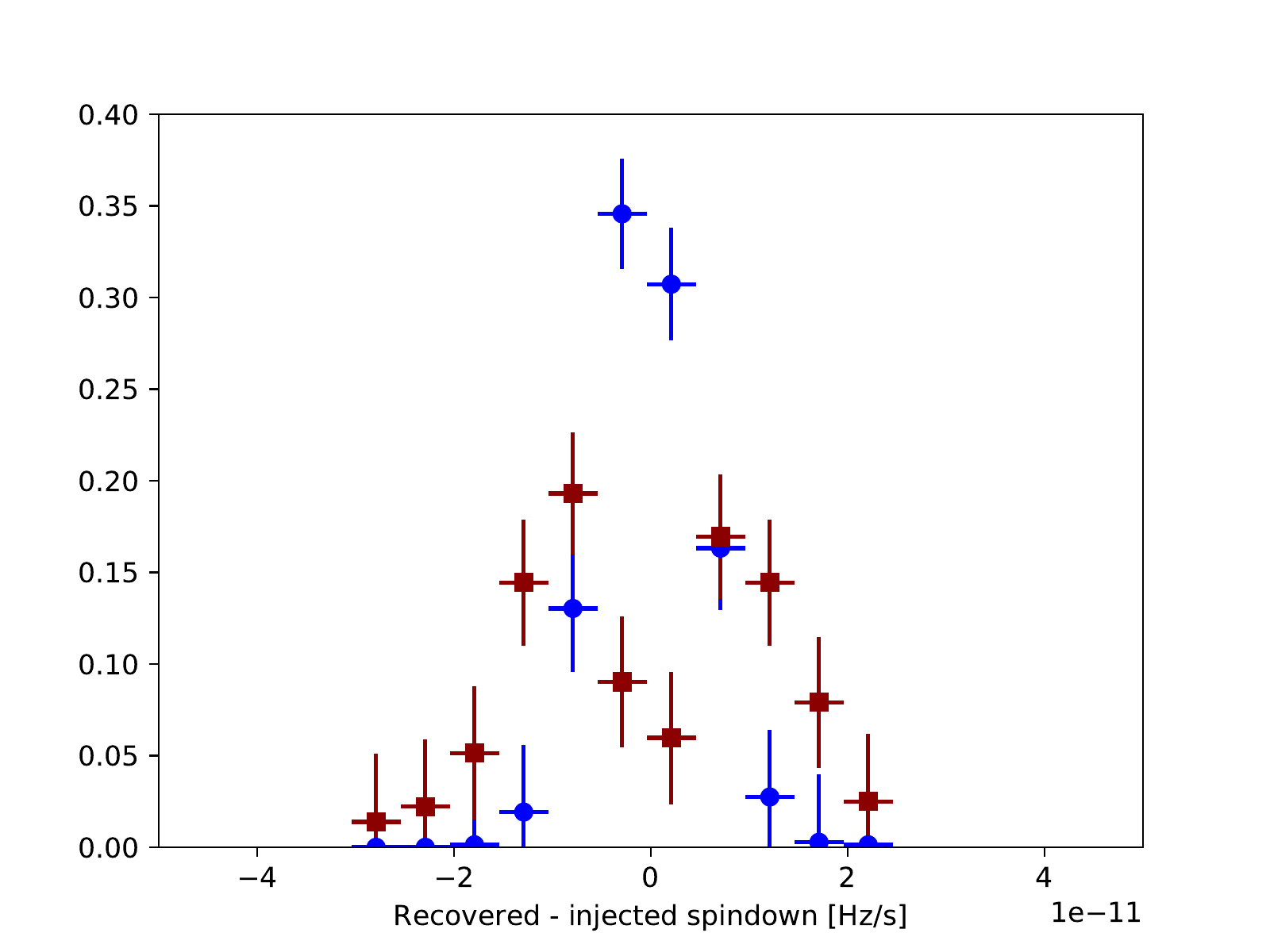}\\
  \includegraphics[width=3.2in]{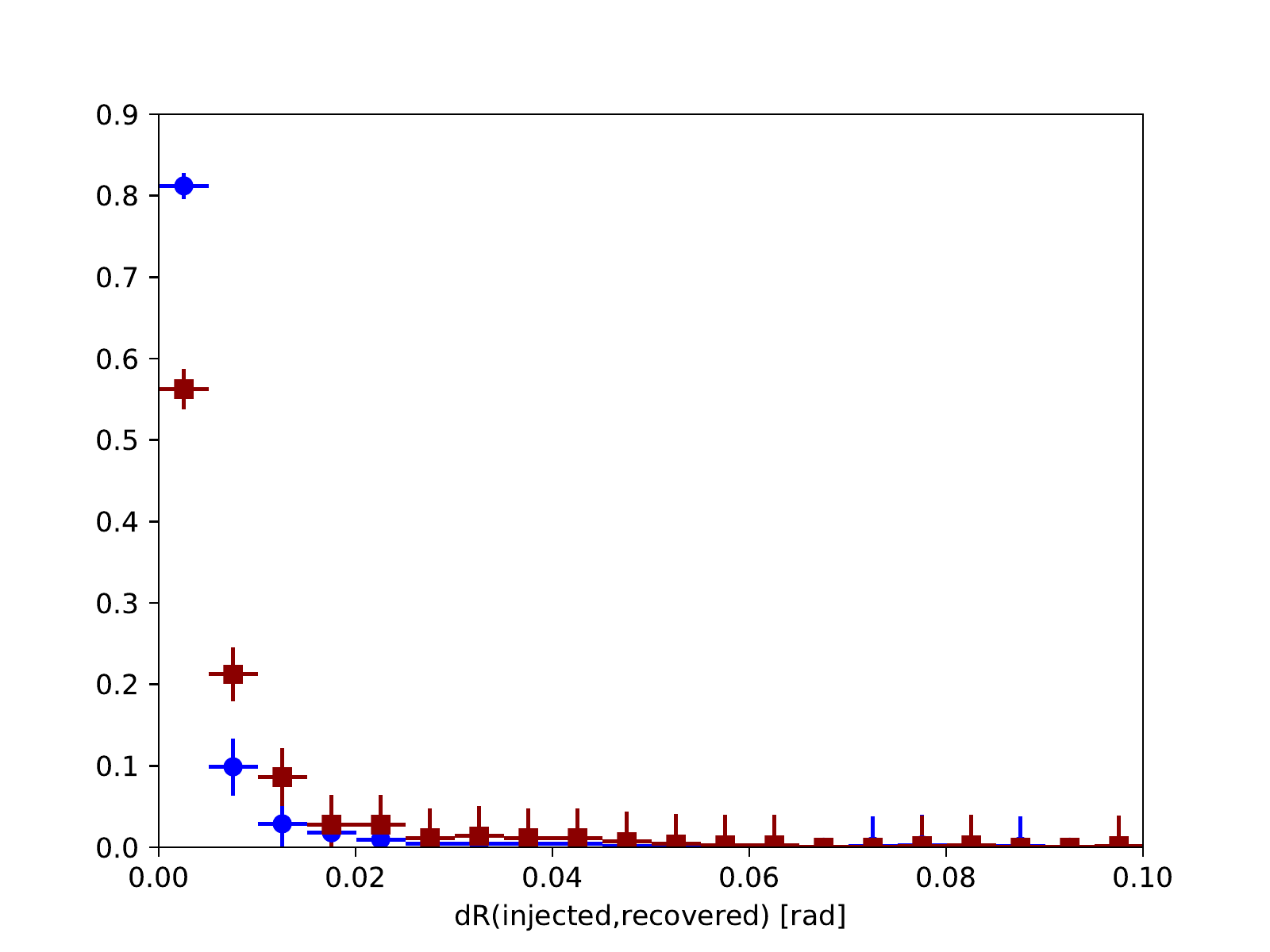}\\

\caption{\label{fig:distances_merged_hist} The distance between the signal and the recovered candidate, in frequency, spindown and sky position, when the candidate with the highest SNR is chosen for Weave (blue) and GCT (red). The search can recover more than one candidate per injection. For strong signals, there are in fact many detection candidates around the signals true parameter values. We study the distribution for the distances of candidates from the true signal parameter values for the detection candidates with the highest $2\mathcal{\overline{F}}$. The dR distribution extends beyond the limits shown in the plot with the max dR value at 0.27 for Weave and 0.38 for GCT.}
\end{figure}

The distance between the signal parameters and the parameters recovered in the MDC, in frequency, spindown and sky position, is shown in Figure \ref{fig:distances_merged_hist}. Here the distance in sky is represented by dR, the angular separation between two sky positions in radians. The dR scales approximately proportional to the frequency of the signal. Here we see that the Weave candidates have a smaller uncertainty on the signal parameters. 

For the GCT search the first refinement stage searches $\delta f \pm 1.9\ee{-4}$\,Hz, $\delta \dot{f} \pm 3.46\ee{-11}$\,Hz/s and a sky patch of 1.2 initial-search sky grid bins around the selected candidate. After the GCT search $\sim 90\%$ of signals have the loudest candidate within this region around the signal parameters. We find that after the Weave search $\sim 98\%$ of the loudest candidates recovered are within this region around the signal parameters.

After the Weave search, $ 90\%$ of signals have the loudest candidate within $\delta f \pm 3.9 \ee{-5}$\,Hz, $\delta \dot{f}\pm 8.2 \ee{-12}$\,Hz/s and a sky patch with a radius 0.3 the radius of the GCT $ 90\%$ confidence region. The computing power used to follow-up these candidates scales roughly linearly with $\delta f$, $\delta \dot{f}$ and the square of the sky patch radius. Per candidate, the first Weave refinement stage would be $\sim 200$ times less computationally expensive than the GCT first refinement stage. This means that for the same computational cost, we can reduce the $2\mathcal{\overline{F}}$ threshold to follow-up $200$ times more candidates.  

Alternatively, we can increase the size of the first refinement stage search so that more signal is recovered by the refinement. We find that after the Weave search $ 97\%$ of signals have the loudest candidate within $\delta f \pm 6.3 \ee{-5}$\,Hz, $\delta \dot{f} \pm 1 \ee{-11}$\,Hz/s and a sky patch with a radius 0.4 the radius of the GCT $ 90\%$ confidence region. Per candidate, the first Weave refinement stage is $\sim 70$ times less computationally expensive than the GCT first refinement stage. This means for the same computational cost, we can reduce the $2\mathcal{\overline{F}}$ threshold to follow-up $70$ times more candidates.  

In an all-sky search with Weave, with $N_{eff} = 0.69 N_{\mathrm{templates}}$, a threshold of $2\mathcal{\overline{F}} > 6.155$ would produce 16 million false alarms in Gaussian noise. With a $\sim 70$ times faster follow-up search per candidate, we can afford to followup 1120 million candidates, corresponding to a $2\mathcal{\overline{F}}$ threshold of 5.91. Decreasing $2\mathcal{\overline{F}}_\mathrm{threshold}$ from 6.155 to 5.91 opens the possibility of detecting a GW strain which is $\approx 6\%$ smaller, as the minimum detectable $h_0^2$ is proportional to the non-centrality parameter, $2\mathcal{\overline{F}} - 4$. The $2\mathcal{\overline{F}}$ thresholds for the comparison in this paper are shown in Table \ref{tab:min2F}.

\begin{table}[h!]
\begin{tabular}{|c|c|}
\hline
Search &$2\mathcal{\overline{F}}_\mathrm{threshold}$\\
\hline
GCT & 6.17 \\
Weave & 5.91 \\
\hline
\end{tabular}
\caption{A signal candidate must have $2\mathcal{\overline{F}} > 2\mathcal{\overline{F}}_\mathrm{threshold}$ to be passed to the first refinement stage. The GCT threshold is the same as used for the MDC \cite{MDC}. The Weave threshold was derived in Sections \ref{sec:effective_ntemplates} and \ref{sec:parameter_estimation}.}
\label{tab:min2F}
\end{table}

\subsubsection{MDC detection efficiency}
\label{sec:MDC_efficiency}

We are primarily concerned with the ability of the searches to recover CW signals. The detection efficiency is the fraction of signals which are considered detected, and it is the benchmark that we will use to compare pipeline performance. The detection efficiency is measured as a function of the inverse of signal strength, $h_0$, normalised to the noise of the detector $\sqrt{S_{h}}$.

\begin{figure}[htb!]
  \includegraphics[width=3.5in]{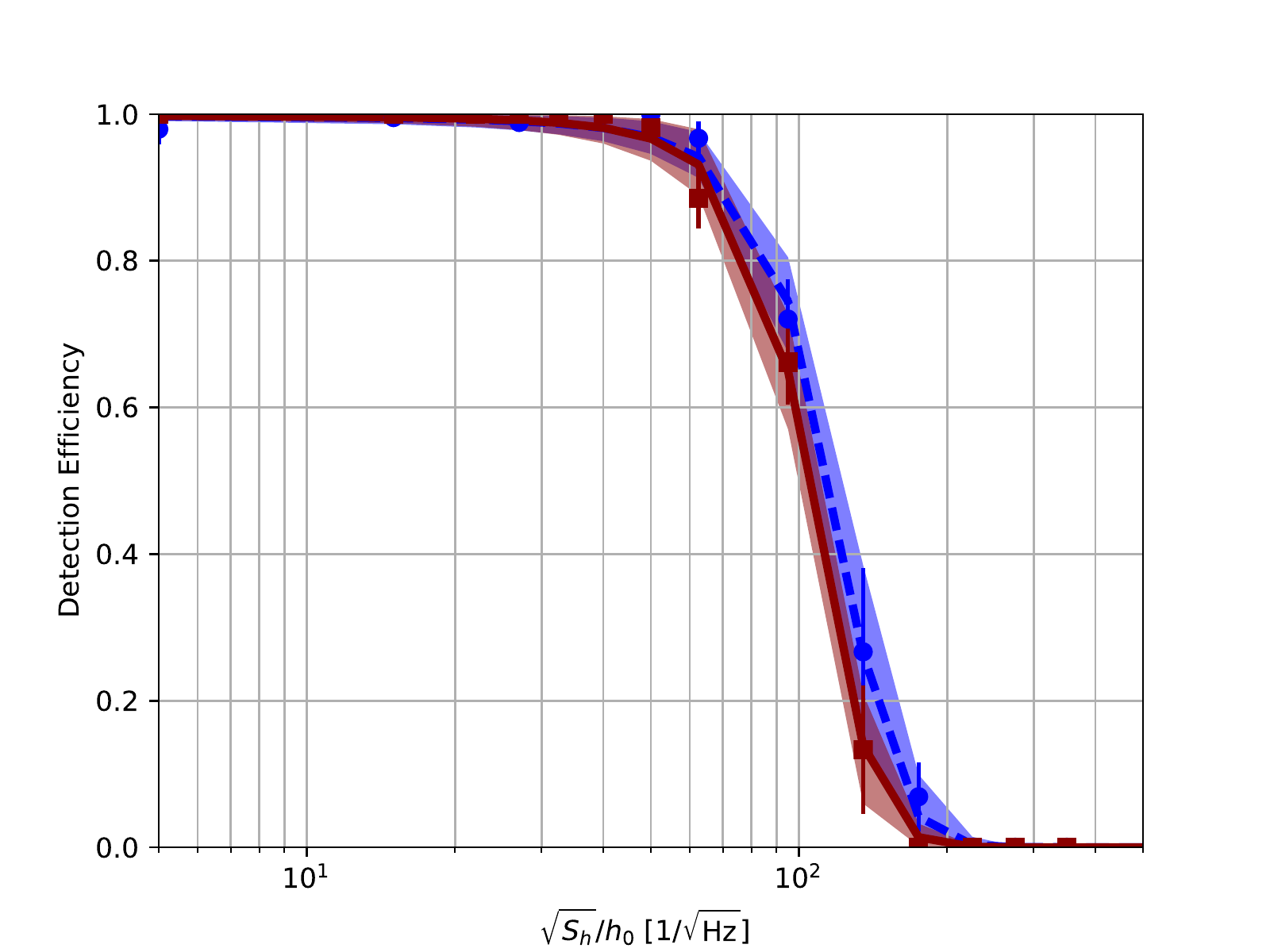}\\
\caption{\label{fig:MDC} Measured detection efficiency of Weave (blue dashed line) and GCT (red solid line) for the MDC injections. The curves and error bands are obtained by fitting sigmoids to the data, see Section \ref{sec:MDC_efficiency}. Note that in \cite{MDC} the signal strength is normalised by a different definition of $S_h$, where the harmonic sum over per-detector PSDs is used instead of the harmonic mean. One can approximately convert the MDC results into sensitivity depth at any efficiency by multiplying by $\sqrt{N_\mathrm{det}} \approx 1.4$, as 2 detectors were used.}

\end{figure}

Unlike \cite{MDC}, we only consider injections which do not overlap with known detector artefacts. For this comparison we are only concerned with the ability of the Weave search method to recover signal in well behaved data. The detection efficiency, measured for MDC injections which do not overlap with a known noise line, is shown in Figure \ref{fig:MDC}. This can be compared directly with Figure 3 in \cite{MDC}.

In Figure \ref{fig:MDC}, Weave appears to be more sensitive than GCT. However, the uncertainty region is too large to quantify the improvement. In the next section we compare the detection efficiency with the new set of fake signals described in Section \ref{sec:MDC}.

The dependence of the detection efficiency on signal strength is obtained with a sigmoidal fit to the MDC results. A non-linear least squares fit is performed using the \textit{scipy.optimize.curve\_fit} package in python, in the same way as done for the MDC. 

\subsubsection{Detection efficiency in Gaussian noise}
\label{sec:res_detection_efficiency}

In this section we compare the detection efficiency of the Weave and GCT search methods using signals in Gaussian noise. The signal parameters are described in Section \ref{sec:MDC}. The main difference with the MDC signals is that these signal are chosen to be randomly uniform in $\sqrt{S_{h}}/h_0\, (\frac{1}{\sqrt{\mathrm{Hz}}})$, so that at any fixed $\sqrt{S_{h}}/h_0\, (\frac{1}{\sqrt{\mathrm{Hz}}})$ the distribution of cos $\iota$ is uniform. This results in a lower apparent efficiency than measured in Figure \ref{fig:MDC}, where the signals were chosen to be randomly uniform in SNR, so that at a fixed $\sqrt{S_{h}}/h_0\, (\frac{1}{\sqrt{\mathrm{Hz}}})$ there are more signals at higher $|$cos $\iota |$.

While the distribution of detection efficiency in Figure \ref{fig:MDC} appears to have a sigmoid shape, we find that when we increase the precision of our efficiency measurement, the distribution is not well described by a sigmoid. This has also been discussed in \cite{Sensitivity2018}. To estimate the sensitivity depth at 90\% detection efficiency we perform a linear fit between the surrounding data points, again, by performing a non-linear least squares fit using the \textit{scipy.optimize.curve\_fit} package in python. The uncertainty band is obtained with a linear fit to the binomial uncertainty on the measured efficiency. This straight line is not expected to desribe the measured efficiency across a broad range of sensitivity depths. 

At 90\% detection efficiency, Weave is 14\% more sensitive than GCT with a sensitivity depth of $57.9\pm0.6$ $\frac{1}{\sqrt{\mathrm{Hz}}}$ at 90\% detection efficiency, compared to $50.8^{+0.7}_{-1.1}$ $\frac{1}{\sqrt{\mathrm{Hz}}}$ for GCT. If we assume CW sources are isotropically distributed, this corresponds to a 50\% increase in the volume of sky where we are sensitive. 

\begin{figure}[htb!]
  \includegraphics[width=3.5in]{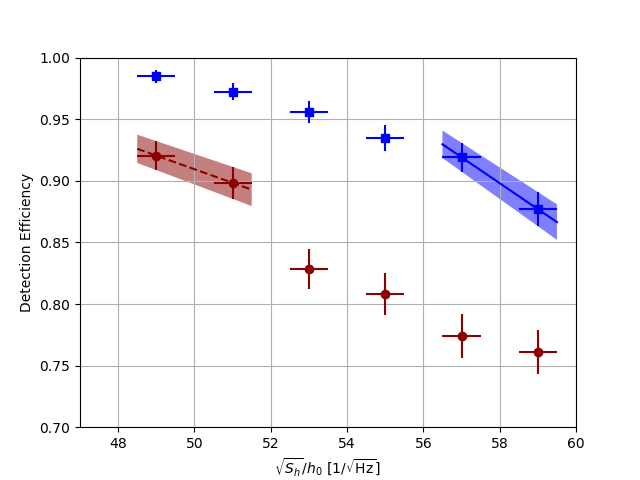}\\
\caption{\label{fig:zoom} Measured detection efficiency for Weave (blue dashed line) and GCT (red solid line) for injections in Gaussian noise which are uniform in $|$cos $\iota |$ at fixed $\sqrt{S_{h}}/h_0\, (\frac{1}{\sqrt{\mathrm{Hz}}})$.}
\end{figure}

\section{Hierarchical Follow-up of interesing candidates}

As mentioned in Section \ref{sec:parameter_estimation}, above-threshold candidates from an all-sky search are passed through hierarchical refinement stages designed to exclude candidates from noise while retaining signal candidates \cite{S6FU,EaH_O1}. At each refinement stage, a search is performed in the parameter space around the candidate with a higher coherent segment length and/or a finer search grid than the previous stage. 

At each refinement stage, the uncertainty on the signal parameters from the previous stage determines the region in parameter space which needs to be searched. In Section \ref{sec:parameter_estimation} we show that the Weave search recovers candidates closer to the true signal parameters than the GCT search. With the smaller parameter uncertaintly, we would have more computing power to invest in using a larger segment length and finer grids at each refinement stage. 

We consider the possibility of using the Weave search from Table \ref{tab:setup_MCs}, to follow-up above threshold candidates from an all-sky search with the GCT search from Table \ref{tab:EaH_setup}. Such a follow-up could search, for example, $\pm 2.5\ee{-4} Hz, \pm 4\ee{-11} Hz/s$ and a sky patch with a radius of 2.0 initial-search sky grid bins around each candidate. This is larger than the 90\% confidence region mentioned in \ref{sec:parameter_estimation}.

The runtime for the Weave search over this parameter space is less than 10 seconds, and the memory is 50 Mb. For a computing cluster with 5000 CPUs, it would take 9 hours to follow-up the 16 million candidates from GCT with this Weave search. This small investment would result in candidates recovered much closer to the true signal parameters.

Each stage of the hierarchical refinement procedure requires the user to determine the search setup for that stage. This is done using the same extensitve Monte Carlo simulations used to determine the search setup for the initial all-sky search. 

\section{Conclusion}

We have considered the state-of-the art Weave search method in its current implementation for Einstein@Home all-sky searches for continuous waves from isolated NS. We compare this new method with the GCT search method currently used in Einstein@Home searches. The comparison is initially made by performing the MDC which has previously been used to compare the methods used for all-sky searches for isolated NS in LIGO and Virgo data \cite{MDC}. However, for a more precise comparison we generate a larger set of signals in Gaussian noise.   

We compare the detection efficiency of an Einstein@Home all-sky search using the GCT search method and the new Weave search method. We find that the Weave search achieves 14\% higher sensitivity depth than the GCT search. This increase translates to a 50\% increase in the volume of parameter space within which we are sensitive to a CW signal. 

The Weave search designed here, for 9 months of LIGO S6 data, is also expected to be optimal for an all-sky search over the 9 months of LIGO O2 data, though in O2 there is a 1 month gap which may affect the optimisation procedure.

While we measure an improvement in sensitivity, the Weave method has disadvantages when considered for a search on Einstein@Home. As explained in Section \ref{sec:memory}, the Weave search requires more memory than a GCT search. This memory needs to be well constrained before deploying the search on volunteer computers, some of which have limited memory capabilities. With the Weave search considered here, limiting the memory to something which can be deployed on Einstein@Home results in a 50\% increase in the search runtime. This memory limitation could potentially be mitigated by adapting the Weave software for multi-threading. 

Furthermore (Section \ref{sec:linerobust}), including the $\BSNtsc$-statistic in an all-sky search with Weave would increase the runtime and memory of the search by a factor of 2 to 3. If less than $\approx 5$\% of frequency bands are disturbed, the $\BSNtsc$-statistic could be applied to recover disturbed bands, however, if 10-20\% of the frequency bands contain disturbances, rerunning with the $\BSNtsc$-statistic becomes prohibitively expensive.

A major benefit of the Weave method is the potential to predict the search sensitivity, which should make it possible to semi-analytically determine optimal search setups. Work is ongoing to apply the optimization framework to the Weave method, which would greatly simplify the process of setting up an Einstein@Home search. The Weave memory model could be incorporated into the optimization framework to find a search setup with low memory requirements. In addition, having an semi-analytic optimization framework would allow us to consider a broader range of search options, as we would not be limited by the time and computing cost of extensive injections-and-recovery studies. 

We find that Weave is promising for the hierarchical follow-up of candidates from an all-sky search. These candidates tend to come from regions of parameter space which have been identified as not being disturbed by detector artefacts, so the search can be run without using the $\BSNtsc$-statistic. These follow-ups tend to run on clusters, so can handle larger memory footprints. However, since these searches cover a small region of parameter space around each candidate, the Weave memory requirement is small. Since the signal parameter recovery is more accurate with Weave, Section \ref{sec:parameter_estimation}, each follow-up stage would cover a smaller parameter space than if the candidates were from a GCT search. This allows for more sensitive follow-up searches at each stage than would be possible with a GCT search. Finally, we need to perform injection-and-recovery searches to determine the search setup to use for each stage. When the semi-analytic optimization framework can be applied to determine the best Weave search setup without injection-and-recovery studies the design of hierarchical follow-up searches will be greatly simplified.

We conclude that the GCT method remains the best search method for deployment on the Einstein@Home computing project, due to its lower memory requirements, while Weave presents significant advantages for the subsequent hierarchical follow-up searches of interesting candidates.

\acknowledgments

The authors would like to thank Heinz-Bernd Eggenstein for useful insights, and for improvements to software used in this paper. Simulations were performed on the Atlas computer cluster of the Max Planck Institute for Gravitational Physics. SW is supported by the National Science Foundation grant NSF PHY 1607585. KW is supported by Australian Research Council grant CE170100004.

\end{document}